\documentclass[preprint,10pt,twocolumn,tightenlines,prd,amsmath,amsfonts,amssymb,nofootinbib,showpacs,floatfix,superscriptaddress]{revtex4}

\usepackage{epsfig}

\newcommand{\gsim}{\buildrel > \over {_\sim}}
\newcommand{\lsim}{\buildrel < \over {_\sim}}
\newcommand{\be}{\begin{equation}}
\newcommand{\ee}{\end{equation}}
\newcommand{\ba}{\begin{eqnarray}}
\newcommand{\ea}{\end{eqnarray}}
\newcommand{\bi}{\begin{itemize}}
\newcommand{\ei}{\end{itemize}}
\newcommand{\nn}{\nonumber}
\newcommand{\eq}[1]{(\ref{#1})}

\newcommand{\ie}{{i.e.}}
\newcommand{\eg}{{e.g.}}

\newcommand{\lqcd}{\Lambda_{QCD}}

\renewcommand{\vec}[1]{\mbox{\boldmath $#1$}}

\newcommand{\qu}{{\rm q}}
\newcommand{\qb}{{\rm\bar q}}
\newcommand{\qq}{\qu\qb }
\newcommand{\QQ}{{\rm Q \bar Q}}

\newcommand{\kvec}{\vec k}
\newcommand{\rvec}{\vec r}
\newcommand{\Rvec}{\vec R}
\newcommand{\x}{x_{B}}
\newcommand{\m}{m_{p}}
\newcommand{\Pom}{\mathbb{P}}
\newcommand{\xP}{x_{\mathbb{P}}}
\newcommand{\ieps}{i\varepsilon}

\begin{document}

\title{Hard Diffraction from Parton Rescattering in QCD}

\author{Stanley J.~Brodsky}
\email[Email:\ ]{sjbth@slac.stanford.edu} 
\affiliation{Stanford Linear Accelerator Center, Stanford CA 94309}

\author{Rikard Enberg}
\email[Email:\ ]{REnberg@lbl.gov} 
\thanks{\\Present address:\ Theoretical Physics Group, Lawrence Berkeley 
National Laboratory, Berkeley, CA 94709, USA}
\affiliation{Centre de Physique Th\'eorique, \'Ecole Polytechnique, 91128 Palaiseau Cedex, France}

\author{Paul Hoyer}
\email[Email:\ ]{paul.hoyer@helsinki.fi} 
\affiliation{ Department of Physical Sciences and Helsinki Institute of Physics\\
POB 64, FIN-00014 University of
Helsinki, Finland}

\author{Gunnar Ingelman}
\email[Email:\ ]{gunnar.ingelman@tsl.uu.se} 
\affiliation{High Energy Physics, Uppsala University, Box
535, 751~21 Uppsala, Sweden}
\affiliation{DESY, Notkestrasse 85, 22603 Hamburg, Germany \\ ~ }

\begin{abstract}
We analyze the QCD dynamics of diffractive deep
inelastic scattering. The presence of a rapidity gap between the target and diffractive system
requires that the target remnant emerges in a color singlet state, which we show is made possible by the soft rescattering of the struck quark. This rescattering is described by the path-ordered exponential (Wilson line) in the expression for
the parton distribution function of the target. The multiple 
scattering of the struck parton via instantaneous interactions in the target generates 
dominantly imaginary diffractive amplitudes, giving rise to an 
`effective pomeron' exchange. The pomeron 
is not an intrinsic part of the proton
but a dynamical effect of the interaction. This
picture also applies to diffraction in hadron-initiated processes.
Due to the different color environment the rescattering is different in
virtual photon- and hadron-induced processes, explaining the 
observed non-universality of diffractive parton distributions.  
This framework provides a theoretical basis for the phenomenologically
successful Soft Color Interaction model which includes
rescattering effects and thus generates a variety of final states 
with rapidity gaps.
We discuss developments of the SCI model to account for the color coherence features of the underlying subprocesses.
\end{abstract}

\pacs{12.38.-t, 13.60.-r, 13.85.-t, 13.90.+i}

\vspace{-20cm}
\phantom{0}    
\phantom{0}    
\phantom{0}    
\phantom{0}    

\hfill SLAC--PUB--10692 

\hfill CPHT--RR 049.0804 

\hfill HIP--2004--37/TH

\hfill TSL/ISV--2004--0281

\maketitle

\section{Introduction}

Hard diffraction has been an active field of research for 20
years~\cite{IS, UA8,GIreview,AHreview,WMreview}, but 
its dynamics is still not adequately understood from the basic
perspective of quantum chromodynamics.
In diffractive events a high energy
(target or projectile) proton survives the
collision intact and keeps most of its initial
energy, leaving a large gap in rapidity relative to the other
produced particles.  The same phenomenon appears also in hard hadron
collisions producing jets or
weak bosons. The phenomenon of hard diffraction was discovered and first
studied by the UA8 experiment \cite{UA8} in $p\bar{p}$ collisions.
With the subsequent discovery~\cite{ddisexp,Abramowicz:1999eq} of
events with a large rapidity gap in deep inelastic lepton
scattering (DIS), a new window on diffractive dynamics was opened. 
The presence of a hard scale in these processes
suggests that  it should be possible to study hard diffraction
using the analysis tools of QCD perturbation theory.

In the intuitive picture of inclusive DIS a color string-field
is formed between the struck parton and the target remnant
covering the whole rapidity interval between them.  The breaking
of the string during the hadronization process fills the rapidity
interval with hadrons; the probability of a rapidity gap
would be expected to decrease exponentially with the gap size.
However, HERA measurements~\cite{ddisexp,Abramowicz:1999eq} 
show that about 10\% of DIS final
states have a large gap.  Remarkably, in most of these diffractive
deep inelastic scattering (DDIS) events, the target proton
scatters elastically and remains intact. 
The DDIS to DIS cross section ratio depends weakly on the virtuality
$Q^2$ of the photon at fixed Bjorken $x_B$, implying that DDIS is a
Bjorken-scaling leading-twist process.  The inclusive structure
functions determined experimentally from the total DIS cross
section thus have a significant diffractive contribution.

Well before the experimental discoveries, Ingelman and Schlein
(IS)~\cite{IS} proposed hard diffraction, and in particular
diffractive DIS, as a tool for studying the mechanisms underlying
diffractive interactions.  In Regge language, diffraction and the
rapidity gaps which persist at high energy are associated with 
pomeron exchange;  the structure of the `pomeron' could then be
elucidated. If the pomeron could be regarded as a
color-singlet hadronic component within the target proton carrying
a small fraction $\xP$ of the proton momentum, the virtual
photon would probe the quark content of the pomeron itself.  Such
events would have a rapidity gap of size $\sim \log(1/\xP)$ between
the proton and the remaining hadronic final state.  The IS
model, using diffractive parton distributions combined with
next-to-leading order QCD evolution, provides good fits to HERA
measurements~\cite{ddisexp,Abramowicz:1999eq,Adloff:1997sc,Breitweg:1998gc}
of diffractive deep inelastic scattering.

In the IS approach the parameters describing the structure of the pomeron are fixed by
DDIS data. The pomeron structure function should
be universal if the pomeron were indeed an intrinsic 
part of the target proton wave function.
In hadron collisions quarks and gluons of one
hadron may then scatter off constituents of the pomeron in the other
hadron, giving rise to hard scattering events with rapidity gaps.
The initial observations by UA8 \cite{UA8} of jets in diffractive
events agreed qualitatively with this scenario.  However, later
$p\bar p$ collider data from the CDF~\cite{CDF-JJ,CDF-DPE} and
D\O{}~\cite{D0-JJ} collaborations showed that hard-diffractive
events with a pair of high-$p_\perp$ jets constitute
only 1--2\% of all
jet events.  Similarly, in events with $W$ and $Z$ boson
production, only a small fraction 
$\sim 1\% 
$ of events have
rapidity gaps \cite{CDF-W,D0-WZ}.  Thus the diffractive fraction
is observed to be considerably smaller in hadronically-induced
events compared to the DDIS/DIS ratio of about 10\%.

Thus one needs to question the approach of 
expressing the diffractive cross section as a product of a pomeron 
flux from the proton times a universal distribution of partons in the pomeron.
Studies of perturbative models~\cite{Collins:cv,Berera:1994xh} have indeed shown that the QCD factorization theorems~\cite{factorization} do not apply to hard diffractive hadron--hadron scattering. However, diffractive processes {\em induced by virtual photons} such as DDIS do factorize as a product of diffractive parton distributions times the usual hard parton cross sections~\cite{diff-factorization}. 

DDIS models based on a two-gluon exchange picture of the pomeron~\cite{WMreview,Bartels:1998ea}  have been formulated in the target rest frame with the virtual photon splitting into a color dipole at a `Ioffe' distance~\cite{Ioffe:1969kf} $L_I \sim 1/m_px_B$ before the target. The dipole is assumed to consist dominantly of a quark-antiquark pair for small diffractive masses $M_X$, or a
quark-antiquark pair plus one or more gluons for large diffractive masses $M_X$.
The amplitude for the quasi-elastic dipole-proton scattering was evaluated using a gluon distribution of the proton at the target vertex. These models have been successful in describing DDIS data, but it is not clear if and how they can be applied to diffraction in hard hadron--hadron collisions.

Here we shall discuss two other approaches which we will argue actually
represent the same physics---the Soft Color Interaction
(SCI) model originally proposed by Edin, Ingelman and
Rathsman~\cite{SCI} and the framework for soft rescattering in
deep inelastic scattering proposed by Brodsky, Hoyer, Marchal,
Peign\'e and Sannino (BHMPS)~\cite{BHMPS}.  
In both analyses, diffraction is the result of soft rescattering of partons involved in the hard process.

In the conventional parton model picture of DIS the virtual photon is absorbed on a quark in the target. The struck quark then propagates through the target with (nearly) the velocity of light and may interact with the target spectators via longitudinal ($A^+$) gluon exchange. This soft rescattering is described (in a general gauge) by the path-ordered exponential (Wilson line) in the expression for the parton distributions given by the QCD factorization theorems~\cite{factorization}. If the photon momentum is chosen to be in the negative $z$-direction the rescatterings occur (in the Bjorken limit) at an instant of Light-Front (LF) time $x^+ = t+z$. Hence the rescattering can formally be included in the definition of the $x^+=0$ target LF wave function, even though the exchanges occur a finite ordinary time $t \simeq -z$ {\em after} the hard virtual photon interaction. We note that ``augmented'' wave functions defined in this way have counterintuitive properties which do not relate to the structure of the target in isolation---in particular, they have absorptive parts arising from the rescattering dynamics.

In the SCI model diffraction arises from soft gluon exchanges between the target spectators and the diffractive (projectile) system which leaves the target in a color-singlet state.  
The color currents induced by the hard virtual photon interaction must therefore be screened before the onset of hadronization. This is achieved by parton rescattering, which in the BHMPS approach occurs via essentially instantaneous gluon exchange analogous to `Coulomb' scattering.
The rescatterings involve on-shell intermediate states which at small $x_B$
provide the imaginary phase associated with diffractive scattering
or pomeron exchange.\footnote{Although the term \emph{pomeron}
originates in Regge theory, we use it
to generally denote the mechanism which leads to
diffraction without reference to any specific theoretical model or
framework.}
The rescattering is part of the standard leading-twist
DIS dynamics and thus is not power-suppressed at large $Q^2$.
It also causes shadowing effects in nuclear targets~\cite{BHMPS} and
Bjorken-scaling single-spin asymmetries in DIS~\cite{BHS}.

One of our  objectives in this paper is to demonstrate the
interrelationship between the SCI and BHMPS approaches  and to
show  how they describe the same physics.  The SCI model has been
very successful phenomenologically, being able to reproduce the
main features of all data on hard diffraction in DIS and in
hadron--hadron collisions.  It lacks, however, an explicit theoretical
basis and some important features are not accounted for.  These
shortcomings can be improved by the BHMPS picture which is firmly
based on QCD.

\section{Soft rescattering in Deep Inelastic Scattering}

\subsection{Lorentz frame dependence of the description of DIS dynamics} \label{framedep}

Consider the usual inclusive DIS process $\gamma^*(q)+ {\rm p}(p) \to X(p+q)$, where the incoming proton and virtual photon have momenta $p$ and $q$, respectively. 
The center of mass energy is $W^2 = (p+q)^2$ and we define the usual variables
\ba Q^2 &=& -q^2\\
\x &=& \frac{Q^2}{2 p\cdot q} \approx \frac{Q^2}{Q^2+W^2}\\
\nu &=& \frac{p\cdot q}{\m}
\ea where $\nu$ is the photon energy in the proton rest frame and $\m$ is the
proton mass.  DIS applies ideally in the Bjorken limit, $Q^2\to\infty$,
$\nu\to\infty$ with $\x = Q^2/2 \m \nu$ fixed.

In the original parton model approach to DIS developed
by Feynman~\cite{Feynman} and by Bjorken and Paschos~\cite{Bjorken:1969ja},
DIS is described in an infinite momentum frame where the proton has very large
momentum and the virtual photon scatters on one of the quarks in
the proton. This approach can be formulated more concisely using
light-front (LF) quantization~\cite{Brodsky:1997de}, with the LF ``time'' defined by $x^+ = t+z$. 

The LF dynamics is invariant under boosts in the $z$-direction. It is thus not necessary that the target proton have large momentum---for simplicity we here choose the target rest frame. However,
for a parton model interpretation it is essential that the virtual photon 
probes the target wave function at an instant of $x^+$ in the Bjorken limit. This is ensured by choosing the photon momentum in the {\em negative} $z$-direction. We define the ``parton model frame'' of DIS as the frame where the target proton is at rest, $p=(m_p,\vec{0})$ and the photon momentum is directed along the negative $z$-direction, $q=(\nu,0,0,-\sqrt{\nu^2+Q^2})$:
\ba \label{lfframe}
\mbox{Parton model frame:} \hspace{-15mm}  & \\
& p^+ = p^- = m_p, \hspace{.5cm}&{\vec p}_\perp = 0 \nn \\ 
& q^+ \simeq -m_px_B, \ \ q^- \simeq 2\nu, \hspace{.5cm} &{\vec q}_\perp = 0
\nn 
\ea
where we have used the notation $q=(q^+,q^-,\vec{q}_\perp)$.

\begin{figure}[tb]
\epsfig{file=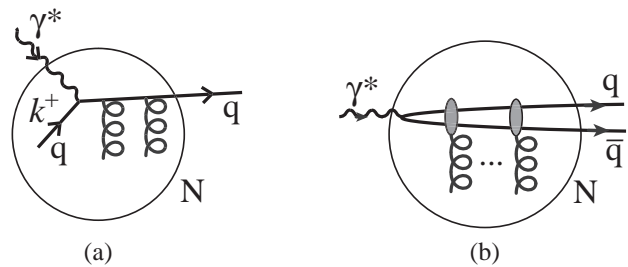,width=0.95\columnwidth}
\caption{Virtual photon scattering on 
the proton in (left) the parton model frame and (right) the dipole frame. The LF time increases from left to right and the rescattering effects are indicated. \label{DIS-vs-dipole}}
\end{figure}

Quarks and gluons propagate forward in LF time $x^+$ if and only if they have positive\footnote{The virtual photon may formally be treated as an `incoming' particle even though it has $q^+ < 0$.} `$+$' momentum. In the parton model frame the virtual photon thus cannot split into a forward moving $\qq$ pair, since this would require $q^+ = p_{\qu}^+ + p_{\qb}^+ > 0$. Instead, the photon scatters off a quark in the target as shown in Fig.~\ref{DIS-vs-dipole}(a). The struck quark absorbs the large `$-$' momentum of the photon and can remain on-shell only if
$p_\qu^+ \simeq (m_\qu^2+p_{\qu \perp}^2)/2\nu \to 0$ in the Bjorken limit. Hence the target quark must have $k^+ = -q^+ = m_px_B = p^+ x_B$, \ie, the target quark carries the LF momentum fraction $x_B$ of the proton momentum. Thus the parton model frame provides the standard `handbag' view of DIS dynamics. The imaginary part of the handbag diagram is given by the probability for finding a quark carrying target momentum fraction $x_B$ (when struck quark rescattering via longitudinal $(A^+)$ gluon exchange is ignored, see section~\ref{rescatsection}). This probability is determined by the target LF wave functions for Fock states defined at equal $x^+$.

While the above parton model frame view of DIS is valid at any $x_B$, it is often helpful to choose another frame when considering small-$x_B$ phenomena such as shadowing and diffraction. At small $x_B$ the target quark (or gluon) has high $k^- \simeq (m_\qu^2+k_\perp^2)/m_px_B$ even before being struck by the photon, and may be treated as a constituent of the photon rather than of the proton. Such a view of DIS dynamics is provided by a frame where the photon moves along the {\em positive} $z$-direction. In this ``dipole frame'' we have then
\ba \label{dipframe}
\mbox{Dipole frame:} \hspace{-5mm}   & \\
& p^+ = p^- = m_p, & {\vec p}_\perp = 0 \nn \\ 
&q^+ \simeq 2\nu, \ \ q^- \simeq -m_px_B, \hspace{.5cm} &{\vec q}_\perp = 0
\nn 
\ea
As the photon propagates through the target, its $z$ coordinate increases with time $t$. Thus the photon does not probe the target at an instant of LF time $x^+$ in the dipole frame, and the DIS cross section is not simply related to the LF wave function of the target. The dipole frame is reached from the parton model frame by a rotation of $180^\circ$ around the $x$ (or $y$) -axis. Such a rotation is a ``dynamical'' Lorentz transformation in the LF formalism (it affects $x^+$); hence relating the two frames requires a complete knowledge of the target wave function.

In the dipole frame $q^+$ is large and the transition $\gamma^* \to \qq$ is allowed, as shown in Fig.~\ref{DIS-vs-dipole}(b). The $180^\circ$ rotation has transformed the incoming target quark of the parton model frame into the $\qb$ of the dipole frame, which now carries large 
$p_\qb^+ \simeq (m_\qu^2+{p_{\qb \perp}}^2)/m_px_B$ at small $x_B$. The $\qq$ pair then forms a color dipole which scatters at high momentum $\sim p_\qb^+$ from the target. 
This view of DIS was suggested long ago in the covariant parton model of Landshoff, Polkinghorne and Short \cite{Landshoff:1971ff}.
As in ordinary hadron scattering one expects the dipole scattering to have a diffractive component mediated by pomeron exchange. The analyses of Refs.~\cite{WMreview,Bartels:1998ea} used the dipole frame with a two-gluon exchange model for the pomeron. Because of the frame choice the relationship of this model to the LF wave function of the target was unclear. Here we shall see that the lower vertex involving the incoming and outgoing proton and two gluons is, in fact, not related to the generalized gluon distribution as conjectured in~\cite{WMreview,Bartels:1998ea}.

\subsection{The soft color interaction model}\label{subsec-sci}

The starting assumption of the SCI phenomenological model is that the underlying hard
interaction of a diffractive event is of the same type as in a
normal event.  This is supported by the similarity of DIS and DDIS
data, \eg\ the flatness of the ratio $\sigma^{\gamma^*
p}_{\text{diff}}/\sigma^{\gamma^* p}_{\text{tot}}$
in both $x_B$ and $Q^2$ \cite{Chekanov:2002bz}.  The color-singlet exchange which leads to rapidity gaps (and leading protons) is caused by 
soft interactions postulated to occur between the partons from the hard interaction and the color field of the proton.

\begin{figure}[t]
\centerline{\epsfig{file=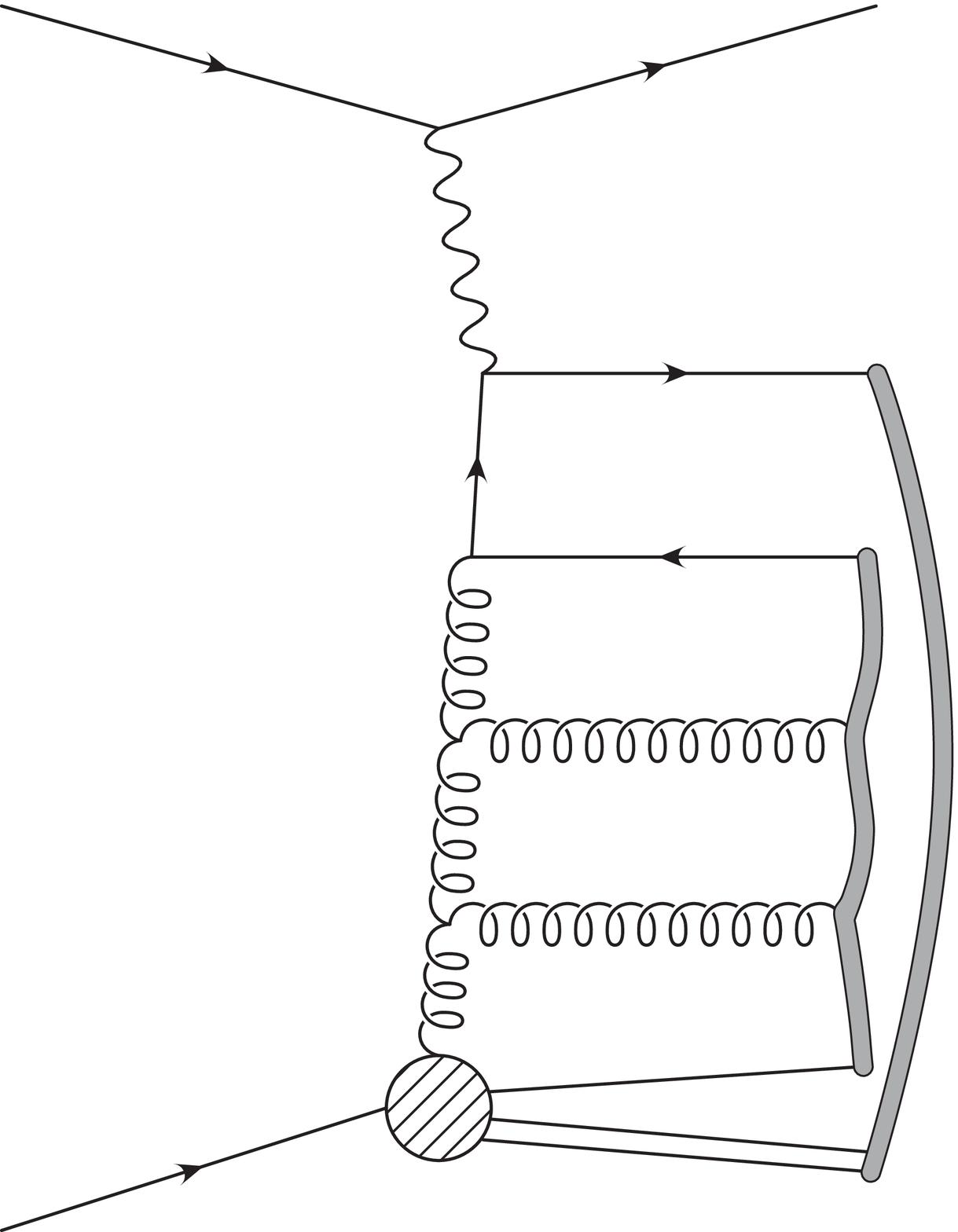,width=0.48\columnwidth}
\epsfig{file=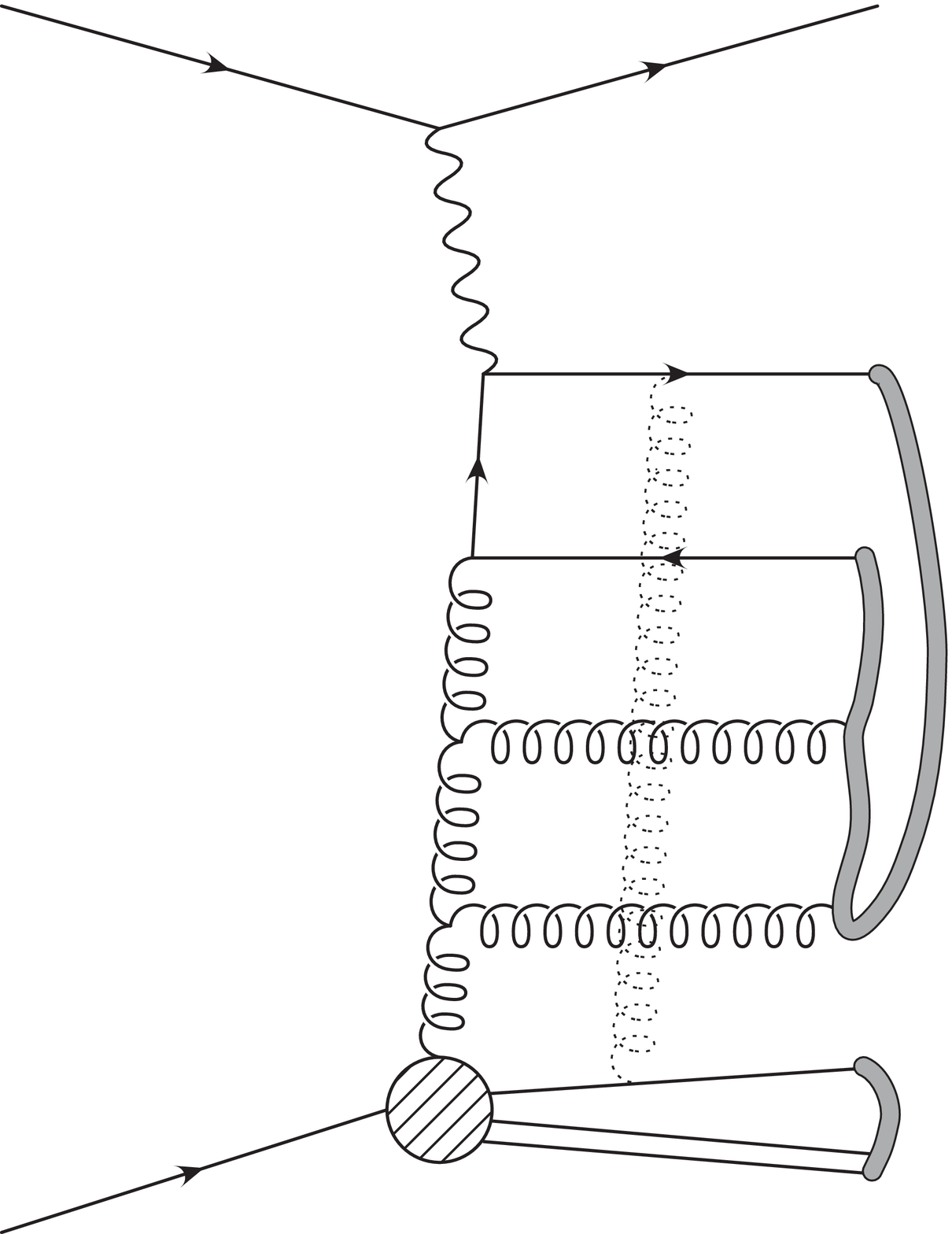,width=0.48\columnwidth}}
\caption{Gluon-induced DIS at small-$x$ with color flux tube, or string, configuration in (left) the conventional Lund string model connection of partons and (right) after a soft color octet exchange (dashed gluon line) between the remnant and the hard scattering system resulting in a phase space region without a string leading to a rapidity gap after hadronization. \label{DIS-strings}}
\end{figure}

The model is implemented in the Monte Carlo event generators 
{\sc Lepto} \cite{lepto} for DIS and {\sc Pythia} \cite{pythia} for hadron--hadron collisions.  The standard treatment of
the hard perturbative dynamics in terms of fixed-order matrix elements and parton showers
based on DGLAP evolution is kept, but the Soft Color Interaction (SCI) model is
then introduced in a subsequent
step before the standard Lund string model \cite{lund} performs
the hadronization process to produce the observable final state.  SCI provides an
explicit model for the rescattering of the emerging hard partons on the color
background field of 
the target proton which is represented by the remnant partons.
These rescattering interactions are modeled as non-perturbative gluon exchanges
with negligible momentum transfer. The gluon color-octet charge implies an
exchange of color between the hard partons and the target remnant.  This in turn
changes the color flow in the event leading to another color string topology, as illustrated in Fig.~\ref{DIS-strings}, and
thereby another hadronic final state after applying the standard hadronization
model.
\begin{figure}[t]
\centerline{\epsfig{file=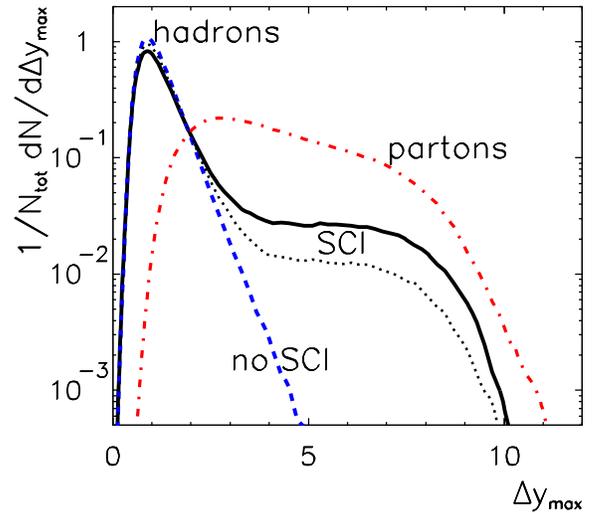,width=1\columnwidth,
bbllx=0, bblly=15, bburx=255, bbury=215}}
\caption{Distribution of the size $\Delta y_{max}$ of the largest rapidity gap in DIS events at HERA simulated using {\sc Lepto} (standard small-$x$ dominated DIS event sample with $Q^2\ge 4$ GeV$^2$ and $x\ge 10^{-4}$).  The dashed-dotted curve represents the parton level obtained from hard, perturbative processes (matrix element plus parton showers). The dashed curve is for the hadronic final state after standard Lund model hadronization, whereas adding the Soft Color Interaction model results in the full curve. The dotted curve is when the SCI probability parameter $P$ has been lowered from its standard value 0.5 to 0.1. \label{gapsize}}
\end{figure}

It is not known how to calculate the probability for such a
non-perturbative gluon exchange.  In the SCI model it is,
therefore, taken as a free parameter $P$ which specifies an
assumed constant probability for each parton to  exchange
color-anticolor (corresponding to a soft gluon) with the remnant
partons.  This parameter is the only new parameter introduced in
the Monte Carlo model in order to account for rapidity gaps.  The value $P\approx 0.5$ is chosen to fit the data on the diffractive
structure function $F_2^D$.  For more details about the model and
the comparisons to data we refer to Refs.~\cite{SCI,Edin:1999jq,SCITEV}.

The soft exchanges of the SCI model thus cause large effects on the
hadronization
process and thereby on final state observables.  This is illustrated in
Fig.~\ref{gapsize}, which shows the largest rapidity gap in Monte Carlo
simulated DIS events at HERA.  
The dramatic difference between the distribution at
the parton level, after all perturbative QCD emissions, and the distribution at the observable
hadron level after hadronization demonstrates that the rapidity
gap observable is highly sensitive to soft dynamics. The introduction of the non-perturbative SCI
mechanism changes the hadron level result drastically, from the dashed to the full curve in Fig.~\ref{gapsize}. In particular, the exponential suppression of large gaps is replaced by an extended $\Delta y_{max}$ distribution, which is characteristic for diffraction. 

The appearance of large rapidity gaps sets in quickly as soon as these soft color exchanges occur, even if  the probability parameter is as small as $P=0.1$, as shown by the dotted curve in Fig.~\ref{gapsize}. Increasing $P$ much above 0.5 leads to a reduction in the gap rate since for more gluon exchanges the color flow may be `switched back' again and no string-free rapidity region be produced. Thus the model is rather stable against variations of the soft color exchange probability.

The SCI model has been successful in reproducing the diffractive data from HERA.
As shown in detail in \cite{Edin:1999jq}, the model can account for the H1 data \cite{Adloff:1997sc} on the diffractive structure function $F_2^D(x_\Pom,\beta,Q^2)$, both in normalization and in dependences on the variables. The description of the data is surprisingly good, in view of the fact that only one parameter is adjusted. Furthermore, the model gives a smooth transition from diffractive to non-diffractive events, as in the data, since it generates events with varying gap size and with no gap at all. 

By moving the SCI program code from {\sc Lepto} to {\sc Pythia} exactly the same model can be applied to hadron--hadron collisions. The available data on hard diffraction from the CDF and D\O\ collaborations at the Tevatron can be described as illustrated in Table~\ref{tab-gapratios} and shown in detail in \cite{SCITEV}. Note that this is achieved with the same value of the single parameter $P$ as obtained from the HERA rapidity gap data. The cross section ratios for single
diffraction are reproduced, as well as the rate for double pomeron exchange measured by CDF~\cite{CDF-DPE} and several kinematical distributions~\cite{SCITEV,SCIGAM}.

\begin{table}[t]
\begin{center}
\begin{tabular}{l@{\hspace{2mm}}ll@{\extracolsep{2mm}}ll}
\hline
\hline
Process     & \multicolumn{2}{l}{Experiment} & \multicolumn{2}{c}{Ratio [\%]} \\
            &      &                 & Observed                & SCI \\
\hline
$W$         & CDF  & \cite{CDF-W}    & $1.15 \pm 0.55$         & 1.2 \\
$Z$         & D\O\ & \cite{D0-WZ}    & $1.44^{+0.62}_{-0.54}$  & 1.0$^\dagger$\\
$b\bar b$   & CDF  & \cite{CDF-B}    & $0.62 \pm 0.25$         & 0.7\\
$J/\psi$    & CDF  & \cite{CDF-JPSI} & $1.45 \pm 0.25$         & 1.4$^\dagger$\\
dijets      & CDF  & \cite{CDF-JJ}   & $0.75 \pm 0.10$         & 0.7 \\
dijets      & D\O\ & \cite{D0-JJ}    & $0.65 \pm 0.04$         & 0.7 \\
\hline
\hline
\multicolumn{5}{l}{$^\dagger$ Predictions made in advance of the data.}
\end{tabular}
\caption{Ratios of diffractive/inclusive for hard scattering
processes  in $p\bar p$ collisions at the Tevatron, showing
experimental results from CDF and D0 compared to the SCI model
calculations of~\protect\cite{SCITEV}.}\label{tab-gapratios}
\end{center}
\end{table}

The SCI model is a simple prescription for Monte Carlo simulations of hard diffractive processes, which was motivated by the striking similarities between inclusive and diffractive data. Here we shall see that the soft color exchanges postulated in the model can in fact be identified with the rescattering of the struck partons expected in QCD. This will also allow to pinpoint some deficiencies of the SCI model, in particular the neglect of color coherence in the soft rescattering of transversally compact clusters of partons.

\subsection{Parton distributions and rescattering} \label{rescatsection}

According to the QCD factorization theorem \cite{factorization}, which is based on the properties of perturbative diagrams at arbitrary orders, the quark distribution of the nucleon is given by the matrix element
\ba
f_{\qu/N}(\x,Q^2) = \frac{1}{8\pi} \int dx^- \exp(-i \x p^+ x^-/2) \nn
\\ \times \,
 \langle N(p)|
\bar\psi(x^-) \gamma^+\, W[x^-;0] \, \psi(0)|N(p)\rangle\  \label{PDFdef} 
\ea
where all fields are evaluated at equal LF time $x^+ = 0$ and small transverse
separation $x_\perp\sim 1/Q$. The Wilson line $W[x^-;0]$, 
\be 
W[x^-;0] = {\rm P}\exp\left[ig\int_0^{x^-} dw^- A^+(w^-) \right] \label{POE} 
\ee 
physically represents rescattering of the struck quark on the target spectators. Only the longitudinal ($A^+$) component appears in the path ordered exponential \eq{POE}. This component has no $x^+$ derivative in the Lagrangian and is therefore ``instantaneous'' in $x^+$. 
Soft transverse ($A^\perp$) gluon exchange does not occur within the coherence length of the virtual photon, $x^- \simeq L_I \sim 1/m_px_B$ in the parton model frame \eq{lfframe} as determined by the Fourier transform in \eq{PDFdef}, and later interactions do not affect the DIS cross section.  This ensures that the DIS cross section is proportional to the nucleon matrix element \eq{PDFdef}; however, as shown in Ref.~\cite{BHMPS} the presence of the Wilson line precludes a probabilistic interpretation of the parton distributions.

The Wilson line reduces to unity in LF gauge, $A^+ = 0$. Hence it is sometimes assumed that the path-ordered exponential is just a gauge artifact; \ie, that the $A^+$ gluon exchanges do not affect the DIS cross section at leading twist. This would conflict with our conventional understanding of diffraction and shadowing as arising from the interference of amplitudes with dynamical phases.

This question was studied in some detail in the perturbative model of BHMPS~\cite{BHMPS}. The contribution to the inclusive DIS cross section from the struck quark rescattering indeed vanishes in LF gauge, consistent with the Wilson line reducing to unity. This follows from the form of the LF gluon propagator,
\be \label{lcprop}
d_{LF}^{\mu\nu}(k) = \frac{i}{k^2+\ieps}\left[-g^{\mu\nu}+\frac{n^\mu k^\nu+ k^\mu
n^\nu}{k^+}\right]
\ee
where $n^2=0$ and $n\cdot A = A^+$. The second term is a gauge artifact which cannot contribute to physical amplitudes. In particular, it was seen that the poles at $k^+ = 0$ generated by the propagator \eq{lcprop} are absent from the full BHMPS amplitudes, although they contribute to individual Feynman diagrams. The contributions of the individual diagrams also depend on the $i\epsilon$ prescription used at $k^+ =0$, but the sum of all diagrams is prescription independent.

The BHMPS diagrams with struck quark rescattering vanish individually in the prescription $k^+ \to k^+ -i\epsilon$ due to a cancellation between the two terms in the square brackets of the LF propagator \eq{lcprop}. However, the remaining $k^+=0$ poles in diagrams involving 
interactions within the target spectator system (such as between $p_2$ and $p'$ in Fig.~\ref{DDIS-FSI} below) then give a non-vanishing contribution to the scattering amplitude. In fact their contribution must be equal to that of the struck quark rescattering in Feynman gauge [the first term in \eq{lcprop}], as required by gauge invariance. LF gauge is thus subtle in that 
{\em interactions between spectators contribute to the DIS cross section} through the gauge dependent $k^+=0$ pole terms.

The Wilson line is thus an essential part of the scattering dynamics. It generates phases and interferences between the various rescattering amplitudes, giving rise to observable effects such as leading twist diffraction, nuclear shadowing~\cite{BHMPS}, as well as Bjorken-scaling polarization effects~\cite{BHS}. The Wilson line is similarly important in deeply virtual exclusive processes such as Compton scattering (DVCS) which are given by Generalized Parton Distributions.

\section{Diffraction in Deep Inelastic Scattering}\label{SectionDiffraction}

\begin{figure}[t]
\centerline{\epsfig{file=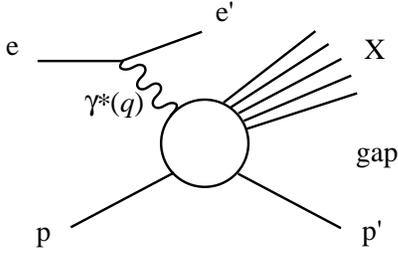,width=0.6\columnwidth}}\caption{Generic diagram for diffractive deep inelastic scattering. \label{DDIS}}
\end{figure}

Let us now consider more closely the diffractive DIS process depicted in
Fig.~\ref{DDIS}. It is convenient to introduce the two additional Lorentz
invariants
\ba \label{vardef}
\xP   &=& q \cdot (p-p^\prime)/q \cdot p  \\
\beta &=&   \frac{Q^2}{2q \cdot (p -p^\prime)}=
 \frac{\x}{\xP} \approx \frac{Q^2}{Q^2+M_X^2} 
\ea 
where $p^\prime$ is the momentum of the final state leading
proton and $M_X$ is the invariant mass of the diffractively
produced system $X$.  The diffractive cross section is then
specified by $(\beta, Q^2, \xP, t)$. In the IS pomeron model~\cite{IS}
$\xP$ is the momentum fraction carried by the pomeron and
$\beta$ plays the role of $\x$ in DIS on the pomeron.  We take the
invariant momentum transfer $t = (p-p')^2$ carried by the pomeron to be small.  
In spite of this conventional
pomeron model interpretation, we stress that these
quantities are defined by the four-vectors of the process and
thus are model independent observables.

\subsection{Mechanism for diffraction}

The perturbative model of DIS studied in Ref.~\cite{BHMPS} provides insights into the dynamics of diffractive DIS and allows one to see why the {\em hard subprocess} is the same as in inclusive DIS, as required by the diffractive factorization theorem~\cite{diff-factorization}. Requiring a rapidity gap between the target and diffractive system imposes a condition only on the {\em soft rescattering} of the struck quark, namely that the target system emerges as a color singlet. As we shall see, this will not modify the $Q^2$-dependence of the cross section.

\begin{figure}[t]
\centerline{\epsfig{file=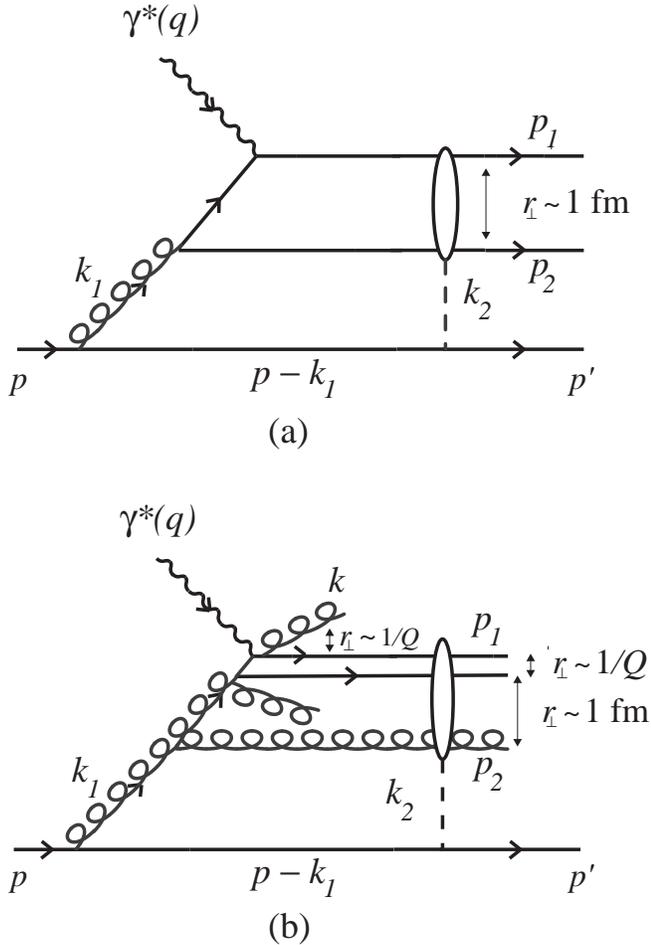,width=1\columnwidth}}
\caption{Low-order rescattering correction to DIS in 
the parton model frame where the virtual photon
momentum is along the negative $z$-axis with $q = (q^+,q^-,\vec
q_\perp) \simeq (-\m \x,2\nu,\vec 0)$ and the
target is at rest, $p=(\m,\m,\vec 0)$. The struck parton absorbs
nearly all the photon momentum, $p_1 \simeq (0,2\nu,\vec
p_{1\perp})$ (aligned jet configuration). In (a) the virtual photon strikes a quark and the diffractive system is formed by the $\qq$ pair ($p_1,p_2$) which rescatters coherently from the target via `instantaneous' longitudinal ($A^+$) gluon exchange with momentum $k_2$. In (b) the $Q\bar Q$ quark pair which is produced in the $\gamma^* g \to Q\bar Q$ subprocess has a small transverse size $r_\perp \sim 1/Q$ and rescatters like a gluon. The diffractive system is then formed by the $(Q\bar Q)\, g$ system. The possibility of hard gluon emission close to the photon vertex is indicated. Such radiation (labeled $k$) emerges at a short transverse distance from the struck parton and is not resolved in the rescattering. \label{DDIS-FSI}} 

\end{figure}

We refer to Ref.~\cite{BHMPS} for a detailed discussion of the properties of the DDIS model amplitudes shown in Fig.~\ref{DDIS-FSI}. Here we only give a qualitative picture of the dynamics of $e p \to e^\prime X p^\prime$ in the parton model frame~\eq{lfframe} as suggested by perturbation theory: 

\vspace{.3cm}\noindent (i) A gluon $(k_1)$ which carries a small fraction $k_1^+/p^+ \sim \xP$ of the proton momentum splits into (a) a $\qq$ or (b) a $gg$ pair. This is a soft process within the target dynamics, consequently the parton pair has a large transverse size $\sim 1$ fm.

\vspace{.2cm}\noindent (ii) The virtual photon is absorbed on (a) one of the quarks in the pair, or (b) scatters via $\gamma^*g \to \QQ$ to a compact, $r_\perp \sim 1/Q$ quark pair. The struck quark (or $\QQ$ pair) carries the asymptotically large photon momentum, $p_1^- \simeq 2\nu$. The parton $(p_2)$ that did not interact with the photon also has large $p_2^- \simeq (m_q^2 + p_{2\perp}^2)/p_2^+$ owing to its small $p_2^+ \sim  x_B p^+$.

\vspace{.2cm}\noindent (iii) Multiple soft longitudinal gluon exchange (labeled $k_2$) turns the color octet $\qq$ of Fig.~\ref{DDIS-FSI}(a) or the $(\QQ) g$ of Fig.~\ref{DDIS-FSI}(b) into a color singlet diffractive system. (The compact $\QQ$ pair behaves as a high energy gluon since its internal structure is not resolved during the soft rescattering.)

\vspace{.3cm}
The rescattering which turns the diffractive system into a color singlet occurs within the target, before it has time to hadronize. The color currents of the gluon exchanges are thus shielded before a color string can form between the target and the diffractive system, hence no hadrons are produced in the rapidity interval $\sim \log(1/x_B)$ between them.

The effective scattering energy of the diffractive system on the target spectator is given by $p_2^- \propto 1/x_B$. As required by analyticity the crossing-even two-gluon exchange amplitude of Fig.~\ref{DDIS-FSI} is imaginary at low $x_B$, implying that the intermediate state between the two gluon exchanges is on-shell. Rescattering is necessary to generate the dominantly imaginary amplitude expected for diffraction.

\begin{figure}[t]
\centerline{\epsfig{file=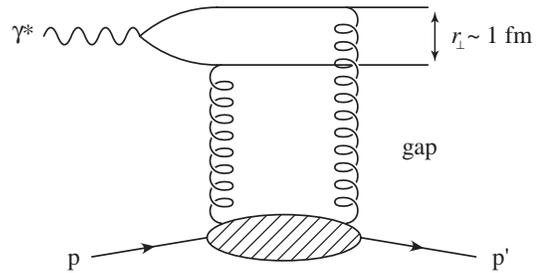, width=0.8\columnwidth}}
\caption{The amplitude of Fig.~\protect\ref{DDIS-FSI}(a) as viewed in the dipole frame. The photon splits into a $\qq$ pair of large transverse size which multiply scatters on the target such that the overall exchange is color neutral.\label{dipolefig}}
\end{figure}

The perturbative amplitude of Fig.~\ref{DDIS-FSI} can also be viewed in the dipole frame~\eq{dipframe}, where the diffractive dynamics appears as shown in Fig.~\ref{dipolefig}. The $\qq$ pair has a large transverse size and is nearly on-shell after the first gluon exchange. The
second gluon represents soft rescattering which occurs in both inclusive and diffractive DIS, and is described by the Wilson line \eq{POE}. Since the upper $\qq$ system is not compact the lower vertex in Fig.~\ref{DDIS-FSI} is unrelated to the generalized parton distribution (contrary to what was assumed in Refs.~\cite{WMreview,Bartels:1998ea}). 

The above scenario involving many soft gluon exchanges is also a feature of the SCI model. The exchanges, which occur in both diffractive and non-diffractive events,
are  assumed to transfer negligible
momentum. However, the SCI is formulated as a Monte Carlo process at the cross section level and hence does not include interference effects.

\subsection{Higher order effects at the hard vertex}\label{evolution}

In the above discussion we have considered the hard virtual photon vertex only at lowest order. Just as in inclusive DIS, hard gluon emission and virtual loops give rise to a scale dependence in the parton distributions, 
and to corrections of higher order in $\alpha_s$ to the subprocess cross section.
In the parton model frame \eq{lfframe} (where the target proton is at rest)
perturbative gluons radiated at the hard vertex in Fig.~\ref{DDIS-FSI}(b)
have $k_\perp \gg p_{2\perp}$ and $k^+ \lsim p_2^+$.
Hence their rapidities $\sim \log(k^-/k_\perp) \sim \log(k_\perp/k^+) \gsim \log(p_{2\perp}/p_2^+)$ tend to be larger than the rapidity of the `slow' parton $p_2$. The hadrons resulting from the hard gluon radiation therefore do not populate the rapidity gap. The gluons are radiated at a short transverse distance from the struck parton and their transverse velocity $v_\perp \sim k_\perp/k^- \sim k^+/k_\perp$ is small. The struck parton and its radiated gluons thus form a transversally compact system whose internal structure is not resolved in the soft rescattering.

According to the above discussion, the size of the rapidity gap and the soft rescattering are unaffected by higher order corrections at the virtual photon vertex. 
This is corroborated by the SCI Monte Carlo, where one observes only small variations of the $\Delta y_{max}$ distribution when varying the parton shower cut-off and thereby the amount of perturbative radiation.
Thus the $Q^2$-dependence of the diffractive parton distributions and the subprocess amplitudes are the same as in inclusive DIS, in accordance with the diffractive factorization theorem~\cite{diff-factorization}.

\subsection{The $x_\Pom$ and $\beta$ dependence of diffractive parton distributions}\label{distshape}

According to the data~\cite{Abramowicz:1999eq,Breitweg:1998gc} the energy ($W$) dependence of diffractive and inclusive DIS are the same within errors. In contrast, in a Regge picture one has
$\sigma_{\text{tot}}^{\gamma^* p} \sim \x^{1-\alpha_\Pom}$
and $\sigma_{\text{diff}}^{\gamma^* p} \sim
\x^{2-2\alpha_\Pom}$.  For a hard pomeron with intercept $\alpha_\Pom
> 1$ the DDIS cross section would then rise faster than the inclusive DIS
cross section as $\x \to 0$.  A similar prediction follows if
the DDIS cross section were given by the square of the gluon
distribution as assumed in the two-gluon exchange model \cite{WMreview,Bartels:1998ea}.

In our picture the underlying hard scattering subprocesses are identical in inclusive and diffractive DIS, involving gluons and sea quarks whose $\x$-dependence reflects the inclusive gluon distribution. The requirement of a rapidity gap places a constraint only on the soft rescattering of the struck parton, namely that the target emerges as a color singlet. The similar $W$-dependence of the DIS and DDIS data thus suggests that little longitudinal momentum is transferred during rescattering. Hence the effective pomeron distribution in the proton should be 
proportional to  
that of the gluon, \ie\ 
\be
f_{\Pom/p}(x_{\Pom}) \propto g(x_{\Pom},Q_0^2) 
\ee 
with $Q_0^2$ the starting scale for the perturbative evolution. 
Furthermore, the quark and gluon structure functions of the pomeron should be similar to the quark and gluon distributions in a gluon, \ie 
\ba \label{pomstrfcn}
f_{q/\Pom}(\beta,Q^2_0) \, & \propto & \, \beta^2+ (1-\beta)^2 \\
f_{g/\Pom}(\beta,Q^2_0) \, & \propto & \, \frac{\left(1-\beta(1-\beta )\right)^2}{\beta(1-\beta )}.
\ea 
These can, however, only be first approximations since using the leading order perturbative splitting functions need not be appropriate for the dominantly small virtuality of the gluon $k_1$ in Fig.~\ref{DDIS-FSI}. 

For diffractive DIS one might at first then guess that the pomeron structure function should be given by the quark component in eq.\ (\ref{pomstrfcn}), corresponding to Fig.~\ref{DDIS-FSI}a. However, the initial $g\to gg$ splitting in Fig.~\ref{DDIS-FSI}b should be important and is likely to dominate at small $x$. The effective pomeron structure function will therefore be a non-trivial combination of the two, which depends on the kinematical variables. For small $\beta$ one may expect $f_{g/\Pom}(\beta,Q^2_0) \, \propto \, 1/\beta$ to give a dominating behavior. Corrections to these expectations may also come from the rescattering gluons, although they are dominantly soft and are effectively included in the parameterizations of the parton distributions of the proton. 
We emphasize again that the pomeron is not an intrinsic part of the proton, but diffraction is a \emph{dynamical effect in the DIS interaction itself.} 

The factorization of the diffractive structure function $F_2^D$ into a pomeron flux and a pomeron parton density is model dependent. We therefore do not pursue the extraction of these pomeron-related functions, but consider instead the model-independent observable $F_2^D(x_\Pom,\beta,Q2)$ which contains the dependence on all variables. As discussed in section \ref{subsec-sci}, the H1 data \cite{Adloff:1997sc} on $F_2^D$ are reproduced by the SCI model \cite{Edin:1999jq}.

\subsection{Color coherence and the SCI model}\label{coherence}

\begin{figure}[t]
\epsfig{file=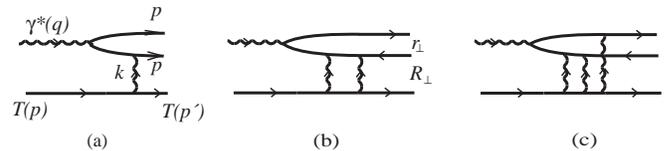, width=\columnwidth}
\caption{A scalar abelian model for deep inelastic scattering with one-, two, and three-gluon exchanges. At each order only one representative diagram is shown. In the $\x \to 0$ limit and at the orders considered no other final states contribute to the total DIS cross section. }
\label{scalardis}
\end{figure}

\begin{figure}[t]
\epsfig{file=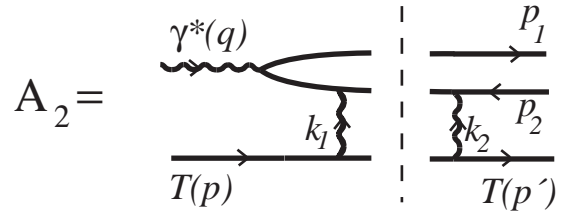}
\caption{The two-gluon exchange amplitude $A_2$ becomes purely imaginary as $\x\to 0$. The intermediate state indicated by the dashed line is thus on-shell and the full amplitude is given by the product of the two subamplitudes on either side of the cut. (Only one of the contributing Feynman diagrams is shown.)}
\label{2gluonfactorization}
\end{figure}

In the above rescattering picture the 
exchanged  
gluons can only resolve the
components 
of the $q-\bar q$ or $q\bar q - g$ system if they have short enough wavelengths, comparable to the transverse size of the dipole. For softer gluons or smaller dipoles there will be destructive interference between diagrams. Expressed differently, the rescattering probability must depend on the transverse momentum of the gluon exchanges.

Let us study this in the toy model presented in \cite{BHMPS}. This is best done in the dipole frame where the transverse impact parameter of the scattering is roughly $1/\sqrt{-t}$ and the size of the dipole is given by the 
relative transverse momentum    
of the quark and antiquark. In this frame it is quite natural to work in the transverse coordinate plane instead of in momentum space.

In \cite{BHMPS}, rescattering in the collision of a virtual photon and a
target 
quark was studied perturbatively up to 3-gluon exchange in scalar Abelian gauge theory, 
see Fig.~\ref{scalardis}. It was found that the one-gluon exchange amplitude in Fig.~\ref{scalardis}(a) is real but the two-gluon exchange amplitude in Fig.~\ref{scalardis}(b) is purely imaginary in the limit $\x\to 0$, and thus the intermediate state between the two gluons is on-shell. Thus the amplitude can be factorized as shown in Fig.~\ref{2gluonfactorization}.

The amplitudes for one-, two-, and $n$-gluon exchange, Fourier transformed to transverse coordinate space, are in this model given by
\begin{align}
A_1 &= {e g^2} \, {\cal C} \, V(m, \x, \rvec_\perp) \, 
W(\rvec_\perp, \Rvec_\perp)  \label{A1} \\
A_2 &= \frac{-i e g^4}{2} {\cal C} \,  V(m, \x, \rvec_\perp) \, 
W^2(\rvec_\perp, \Rvec_\perp) \nonumber \\
&= \frac{-i g^2}{2!} W A_1 \label{A2} \\
&\vdots \nonumber \\
A_n &=  \frac{(-i g^2)^{n-1}}{n!} W^{n-1} A_1 \label{An} 
\end{align}
where ${\cal C}$ is a factor containing kinematical quantities, $V$ is the virtual photon wave function, and $W$ is an eikonal factor given in transverse space by
\be
W(\rvec_\perp, \Rvec_\perp) = \frac{1}{2\pi}
\log\left(\frac{|\Rvec_\perp+\rvec_\perp|}{|\Rvec_\perp|} \right).
\label{W}
\ee 
The transverse distance vector $\rvec_\perp$ is the dipole size and $\Rvec_\perp$ the impact parameter. 
In a numerical simulation these 
can be most simply modeled as the inverses of the corresponding momentum vectors. The 
factors $W$ for each gluon exchange 
arise 
in coordinate space because the vectors $\rvec_\perp, \Rvec_\perp$ are frozen during the scattering process.
It is apparent that the gluon coupling to the dipole decreases when $|\rvec_\perp| / |\Rvec_\perp| \lesssim 1$, which means $|\kvec_\perp| \lesssim 1/|\rvec_\perp|$. 

This effect is not included in the SCI model yet but an obvious method of doing so now suggests itself. It is clear from Eq.~(\ref{A1}--\ref{An}) that additional rescattering happens with a probability amplitude proportional to the factor $W$ in Eq.~(\ref{W}), and that the exchanges can easily be resummed to all orders. This suggests that the SCI model could be modified such that the probability for a soft gluon exchange is not a constant, but is modified by a factor proportional to $W^2$   
for a given $\kvec_\perp$ of the exchanged gluon. The virtuality 
and $\kvec_\perp$ 
of the SCI gluon can be expected to be larger for larger $-t$, so that at large momentum transfer the gluons can couple to smaller dipoles.

The factorization shows that one can include the effect of any number of gluon exchanges in the cross section. This modification would in a natural way include color transparency in the SCI model. 

\subsection{Further tests of color coherence}\label{testcoherence}

A further way to test color coherence aspects may be provided by the dependence of the transverse size of the diffractive $q \bar q$ system on its invariant mass. In the limit of deeply virtual meson production, or of ``exclusive'' jet production as studied in 
\cite{Ashery:2004ta}, the rescattering is hard and the cross section is power suppressed in $Q^2$. The $Q^2$ dependence thus provides a measure of the transverse size of the diffractive system. Increasing the momentum transfer to the target ($t = (p-p')^2$ in Fig. 5) also forces a harder rescattering if the effective target size shrinks with $t$. Correlations between the $t$ and $Q^2$ behaviors thus allow an indirect measurement of the effective target size. This may be phenomenologically investigated in a color coherence version of the SCI model, although the description of more exclusive final states may require other model modifications as well.

\section{Diffraction in Hard Hadron Collisions}\label{SectionHadronic}

Our description of diffraction in deep inelastic lepton scattering can be extended to hard diffractive hadronic collisions. As required by dimensional scaling, only a single parton from the projectile and target participate in the hard subprocess. These leading twist subprocesses (including their higher order corrections) are the same for inclusive and diffractive scattering. The soft rescattering of the hard partons and their spectators is constrained by the requirement of a rapidity gap in the final state. The partonic systems on either side of the gap must be color singlets in order to prevent the formation of a color string in the later hadronization phase.

\begin{figure}[t] 
\centerline{\epsfig{file=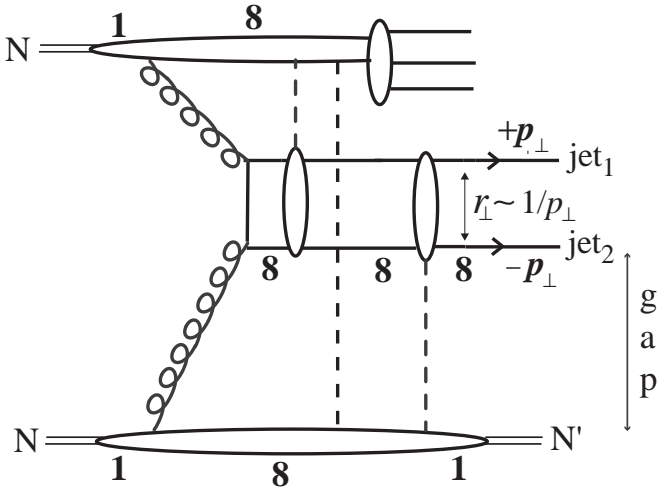, width=1\columnwidth}}
\caption{Illustration of diffraction through rescattering in
$NN \to 2\ jets + X$ in analogy with, and using  
the same notation as, the DIS case in Fig.~\protect\ref{DDIS-FSI}. 
The compact $\qq$ pair which forms the jets is assumed to be in a color octet ({\bf 8}) configuration. This pair rescatters coherently  and thus retains its color.\label{fig-hadronic}}
\end{figure}

The soft rescattering is quite different in hadron collisions as compared to DIS. In hadron collisions both the projectile and target spectator systems are colored. The rescattering gluons ($k_2$ in Fig.~\ref{DDIS-FSI}) can thus couple also to the projectile remnant. In Fig.~\ref{fig-hadronic} the compact $\qq$ pair, which is created in the hard gluon--gluon collision, is not resolved by the soft rescattering and therefore retains it color. Together with the projectile remnant it forms a transversally extended color octet dipole which can rescatter softly from the target remnant. A rapidity gap is formed between the target remnant and the compact $\qq$ pair 
if the target remnant emerges as a color singlet after the rescattering. The probability for this is, however, different from target neutralization in DIS.
This has been previously observed in QCD studies~\cite{Collins:cv,Berera:1994xh} and confirmed experimentally:
the ratio of diffractive to inclusive cross sections is of the order of $\sim 1$\% for a variety of hard processes observed at the Tevatron as compared to the $\sim 10$\% ratio of DDIS/DIS. In the small-$x$ region, these ratios are approximately independent of the momentum fractions in the proton. These observations are well accounted for by the SCI model \cite{SCITEV}.

Events with two gaps and a central dijet system have also been observed at the Tevatron~\cite{CDF-DPE}. These are called double-pomeron exchange (DPE) corresponding to the conventional description where a pomeron is emitted from each of the colliding hadrons, followed by a hard scattering between one parton from
each pomeron.  In our framework these events have soft rescatterings involving both spectator systems, such that the color of both interacting partons is
screened and both spectator systems emerge as color singlets. 
Such events can be generated, \eg, by diagrams like that of Fig.~\ref{fig-hadronic} when the compact $\qq$ pair is created in a color-singlet state and thus does not rescatter in either the projectile or target.
In fact, the SCI model reproduces~\cite{SCITEV} the empirical observations, both in absolute normalization and in
kinematical distributions such as the dijet invariant mass. Again, the underlying hard processes are 
the well-known perturbative QCD subprocesses of fully inclusive cross-sections.
The appearance of one or more rapidity gaps depends on the soft rescatterings which affect the color topology of the event.

\section{Conclusions}

Hard diffractive processes such as diffractive DIS provide new insight into the dynamics of QCD. We have emphasized that the subprocesses with large momentum transfer are universal in all inclusive reactions: they involve a \emph{single constituent} from the projectile and target and are given by perturbative QCD. The parton distributions reflect the LF wave functions of the colliding particles {\em and} the soft rescattering of the partons emerging from the hard subprocess. The rescattering is mediated by longitudinal gluons and occurs `instantaneously' in LF time as the partons pass the spectators. Hard partons which are radiated in the subprocess itself are not resolved by the soft longitudinal gluons which scatter coherently off the color charge of the struck parton. 

In a diffractive process the soft rescattering is constrained by the requirement that the diffractive systems on either side of the rapidity gap are color singlets. Since the configurations of color-charged spectators are different in virtual photon and the various hadron induced diffractive processes, this requirement means that diffractive parton distributions are process dependent. Comparisons of the parton distributions for different projectiles and rapidity gap configurations can thus give valuable information on the rescattering dynamics.

Our description of hard diffractive reactions provides predictions at several levels of accuracy:

\begin{enumerate}
\item The $Q^2$ dependence of all diffractive parton distributions are the same as that of inclusive parton distributions. For DDIS this is a statement of the diffractive factorization theorem \cite{diff-factorization}.

\item  The dependence on the fractional momentum $x$ carried by the parton is similar for diffractive and inclusive distributions. This reflects our assumption that the momentum transferred in the rescattering is small.

\item The dependence on the diffractive mass (or $\beta$) of the diffractive parton distributions arises from the underlying (non-perturbative) $g \to \qq$ and $g \to gg$ splittings  as discussed above.
\end{enumerate}

The Soft Color Interaction model~\cite{SCI} was motivated by the similarity of diffractive and inclusive data and incorporates most of the above features. 
In particular, point 1 and 2
are present in the model since, by construction, the same inclusive parton densities
are always used, and the momentum exchange in the soft interactions can be
neglected. 
There are
close 
similarities but naturally also differences between the SCI model and the rescattering 
theory 
presented here. The main similarities are the timing and the small momenta of the soft exchanges:\ The soft rescattering occurs after the hard vertex (within the coherence length of the hard interaction) but before hadronization of the partonic system. The exchanged gluons therefore have the opportunity to change the color structure of the interaction. They carry small momentum transfer and do not significantly change the momenta of
partons created in  
the hard interaction.

The main difference, on the other hand, is the absence of quantum-mechanical 
coherence  
effects in the SCI model. This means that there is no color transparency in the SCI model---the soft exchanges in the SCI model couple  to compact color singlet objects. Moreover, amplitudes with different numbers of exchanges do not interfere in the SCI model, as they do in the rescattering 
theory. 
This is a general point that is applicable to most Monte Carlo models. Finally, the SCI exchanges carry zero momentum whereas the rescattering gluons have $k^+\sim m\x$ and $k^-, k_T\sim \lqcd$. 

The SCI model may be further developed based on the insights provided above, e.g.\ along the lines discussed in Sect.\ \ref{coherence}.  The SCI model may be seen as a specific model implementation of the general rescattering scenario and some of the differences between the models arise from compromises and assumptions that are necessary to make quantitative comparison with data.
Already in its present form, however, the success of the SCI model in fitting data on diffractive processes in both deep inelastic scattering and hadron collisions lends support to the correctness of the description of diffraction presented here.

\begin{acknowledgments}
We are grateful to St\'ephane Peign\'e and Johan Rathsman for valuable discussions. This research was supported in part by the US
Department of Energy under contract DE--AC02--76SF00515,
by CNRS through Centre de Physique Th\'eorique (UMR 7644 du CNRS),
by the Academy of Finland through grant 102046, and 
by the Swedish Research Council.
RE wishes to thank SPhT CEA-Saclay
for their hospitality when parts of this work was done. 
\end{acknowledgments}


\end{document}